\documentclass[twocol]{ametsoc}

\journal{jas}

\bibpunct{(}{)}{;}{a}{}{,}

\title{The emergence of a summer hemisphere jet in planetary atmospheres}

    \author{Ilai Guendelman\correspondingauthor{Ilai Guendelman, 
     Department of Earth and Planetary Sciences, Weizmann Institute of Science,
234 Herzl st., 76100, Rehovot, Israel}
}

     \affiliation{Department of Earth and Planetary Sciences, Weizmann Institute of Science, Rehovot, Israel}

\email{ilai.guendelman@weizmann.ac.il}

    \extraauthor{Darryn W. Waugh}
    \extraaffil{Department of Earth and Planetary Sciences, Johns Hopkins University, Baltimore, Maryland, USA}
    
    \extraauthor{Yohai Kaspi}
    \extraaffil{Department of Earth and Planetary Sciences, Weizmann Institute of Science, Rehovot, Israel}

\abstract{Zonal jets are common in planetary atmospheres. Their character, structure, and seasonal variability depend on the planetary parameters. During solstice on Earth and Mars, there is a strong westerly jet in the winter hemisphere and weak, low-level westerlies in the ascending regions of the Hadley cell in the summer hemisphere. This summer jet has been less explored in a broad planetary context, both due to the dominance of the winter jet and since the balances controlling it are more complex, and understanding them requires exploring a broader parameter regime. To better understand the jet characteristics on terrestrial planets and the transition between winter- and summer-dominated jet regimes, we explore the jet's dependence on rotation rate and obliquity. Across a significant portion of the parameter space, the dominant jet is in the winter hemisphere, and the summer jet is weaker and restricted to the boundary layer. However, we show that for slow rotation rates and high obliquities, the strongest jet is in the summer rather than the winter hemisphere. Analysis of the summer jet's momentum balance reveals that the balance is not simply cyclostrophic and that both boundary layer drag and vertical advection are essential. At high obliquities and slow rotation rates, the cross-equatorial winter cell is wide and strong. The returning poleward flow in the summer hemisphere is balanced by low-level westerlies through an Ekman balance and momentum is advected upwards close to the ascending branch, resulting in a mid-troposphere summer jet.}

\begin{document}

\maketitle
\section{Introduction}
Zonal jets are ubiquitous features of planetary atmospheres. A common feature of the solar system's terrestrial planetary bodies with a seasonal cycle is a strong jet in the winter hemisphere. Winter jets are observed in Earth's stratosphere and troposphere \citep[e.g.,][]{schneider_2006, waugh2017}, Mars' troposphere \citep[e.g.,][]{waugh_2016}, and Titan's stratosphere \citep[e.g.,][]{flasar2009}. These jets inhibit transport between low and polar latitudes, leading to unique processes in polar regions. For example, polar ozone depletion in Earth's stratosphere \citep[e.g.,][]{schoeberl1991}, condensation and removal of CO$_2$ in the Martian polar atmosphere \citep[e.g.,][]{toigo2017}, and formation of ice clouds in Titan's polar stratosphere \citep[e.g.,][]{DeKok2014}.

\begin{figure}
    \centering
    \includegraphics[width=7.5cm]{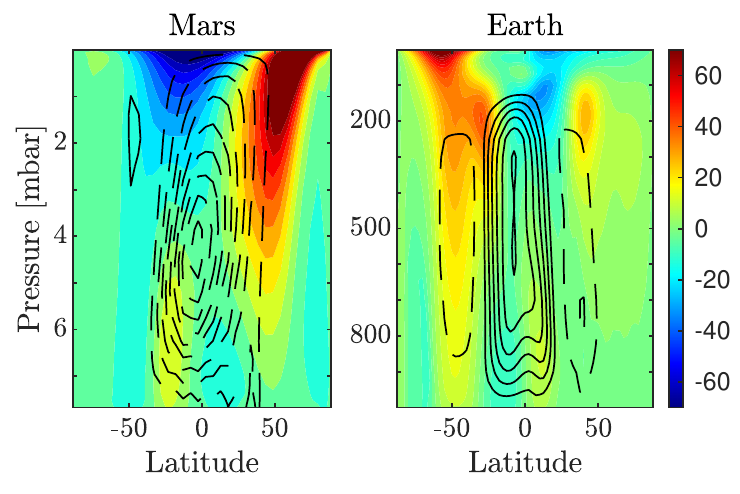}
    \caption{Zonal mean zonal winds (shading) and mean meridional circulation (contours; solid contours are for counter-clockwise circulation, contour intervals are $5\times10^{10}$ and $2\times10^9$ kg s$^{-1}$ for Earth and Mars, respectively). The right panel is of the monsoon region on Earth (Era Interim zonally averaged between $40^{\circ}-100^{\circ}$E); the circulation is calculated from the divergent component of the meridional velocity \citep{raiter2020}, using a June-July-August temporal mean. The left panel is of Mars (EMARS \citep{greybush2019}), using a 90 solar longitude (Ls) mean around Ls 270 (southern hemisphere summer solstice, the season where the low-level jet is strong).}
    \label{fig:obs}
\end{figure}

In addition to the winter jet, low-level westerly winds in the summer hemisphere close to the Hadley cell's ascending edge are also common in other terrestrial planets in the solar system. On Mars, these low-level westerlies are strongest during the southern summer and appear around 30$^{\circ}$S \citep[ e.g.,][see  Fig.~\ref{fig:obs}]{haberle_1993,hinson1999}. On Earth, low-level winds dominate locally in the Indian monsoon region ($\sim$ 15$^{\circ}$N) during the northern summer \citep[e.g.,][see  Fig.~\ref{fig:obs}]{joseph1966existence, findlater1969}. In both cases, the westerlies are close to the surface with easterlies in the upper levels (Fig.~\ref{fig:obs}). Additionally, observations from the waviness of Titan's lake surfaces suggest that low-level winds are stronger during summer \citep{HAYES2013, HOFGARTNER2016}, implying that a similar process might be occurring on Titan. These low-level winds also appear in modeling studies. Specifically, Mars paleoclimate studies have shown that the summer jet's strength and characteristics may have changed during Mars's past, depending on its obliquity \citep[e.g.,][]{haberle_1993,HABERLE2003}.


Although jets are common, their temporal and spatial characteristics vary among planets. This variability is not well understood, requiring a better understanding of the fundamental processes controlling the jet characteristics. In this study, we take an approach of studying these processes in a broad sense rather than focusing on individual models of the specific planets. To better identify the mechanisms responsible for the variability among these planets, we explore, using an idealized three-dimensional general circulation model (GCM) with a seasonal cycle \citep{Guendelman_2019}, a wide parameter space spanning both rotation rate ($\Omega$) and obliquity ($\gamma$). Previous studies have examined the atmospheric dynamics dependence on different planetary and orbital parameters \citep{Williams1982, Williams1988, FERREIRA2014, Mitchell_2014, LINSENMEIER2015, faulk2017, wang2018, guendelman_2018, Guendelman_2019, singh2019, LOBO2020}. However, these studies have either focused on a single parameter, considered equinox conditions, or have not considered the jet structure. 

In section \ref{sec:meth}, we describe the model and the simulations used in this study. In section \ref{sec:flow}, we present the different solstice flow regimes. Surprisingly, we find that for slow rotation rates and high obliquities, the strongest jet is in the summer hemisphere at the ascending branch of a wide, cross-equatorial Hadley cell. To understand the summer jet mechanism, we examine the momentum balances in section \ref{sec:balance}. We show that for the summer jet, the leading order momentum balance cannot be reduced to geostrophic or cyclostrophic, and both the boundary layer drag and the vertical advection play an essential role. Following that, we briefly discuss these regimes in terms of non-dimensional numbers, and we summarize our results in section \ref{sec:conc}.

\section{Model and simulations}\label{sec:meth}

In this study, we use an idealized GCM with a seasonal cycle \citep{Guendelman_2019}. The surface consists of a uniform slab ocean with a constant 10 m depth that exchanges momentum, moisture, and heat with the boundary layer. Moisture effects are represented by simplified condensation and convection schemes \citep{frierson2006}, neglecting the effects of clouds and ice. Radiative processes are accounted for through a two-stream gray radiation scheme, where the longwave optical depth is constant in latitude and varies only with altitude. 

We run simulations in which we vary both the obliquity ($\gamma$), from $10^{\circ}$ to $90^{\circ}$, and the rotation rate ($\Omega$), from $1/32$ to $2$ times Earth's rotation rate, at T42 resolution with 25 vertical layers. We use a $360$-day orbital period with zero eccentricity and otherwise an orbital configuration similar to that of Earth. The model is run for $80$ years, and the climatology is calculated using the latter $50$ years. This study focuses on the solstice response and focuses on the northern hemisphere summer. Results presented here are for a time mean between days $150-240$, equivalent to the June-to-August mean for Earth; the results are insensitive to the period selection.

The model's results with the closest Earth-like simulation ($\gamma=30^{\circ}, \ \Omega=1$, blue highlighted panel in Figure~\ref{fig:jet5x5}) shows good agreement with the observed flow in Earth's troposphere, both in terms of the jet strength and position (Fig.~\ref{fig:obs}). However, there are some differences between the observed and simulated flow, particularly in the stratosphere. This discrepancy is mainly a result of the simplified radiation scheme \cite[e.g.,][]{tan2019} and the lack of an ozone layer in the model.

\section{Flow regimes}\label{sec:flow}

\begin{figure*}[htb!]
    \centering
    \includegraphics[width=16cm]{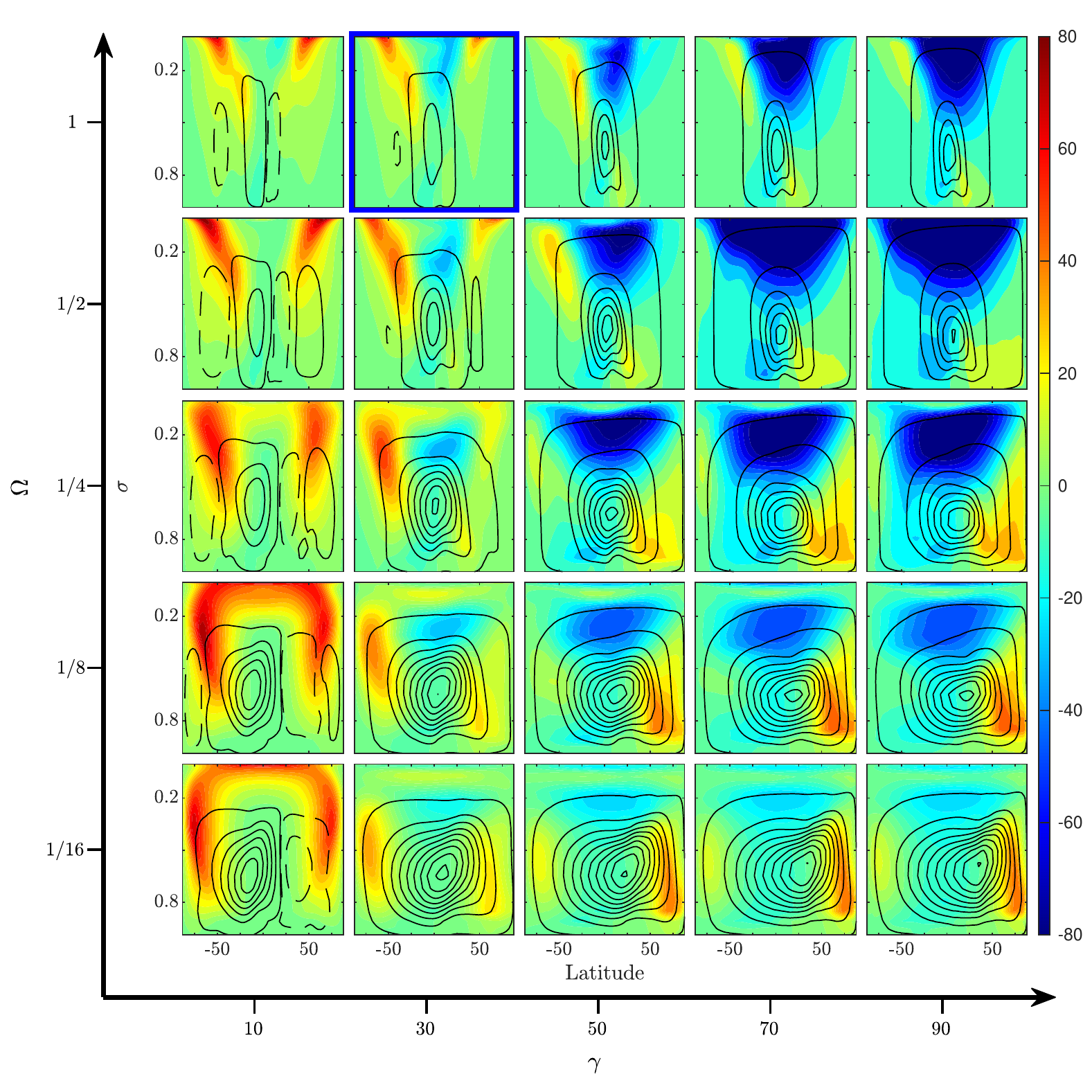}
    \caption{Zonal mean zonal wind (shading, $\pm80$ m s$^{-1}$) and mean meridional circulation (contours, solid contours represent counter-clockwise circulation, contour intervals are $2.5\times10^{11}$ kg s$^{-1}$) for different values of the rotation rate ($\Omega$) and obliquity ($\gamma$). The panel with blue frame represents Earth-like simulations ($\Omega=1$, $\gamma=30^{\circ}$).}
    \label{fig:jet5x5}
\end{figure*}

\begin{figure*}[htb!]
    \centering
    \includegraphics[width=16cm]{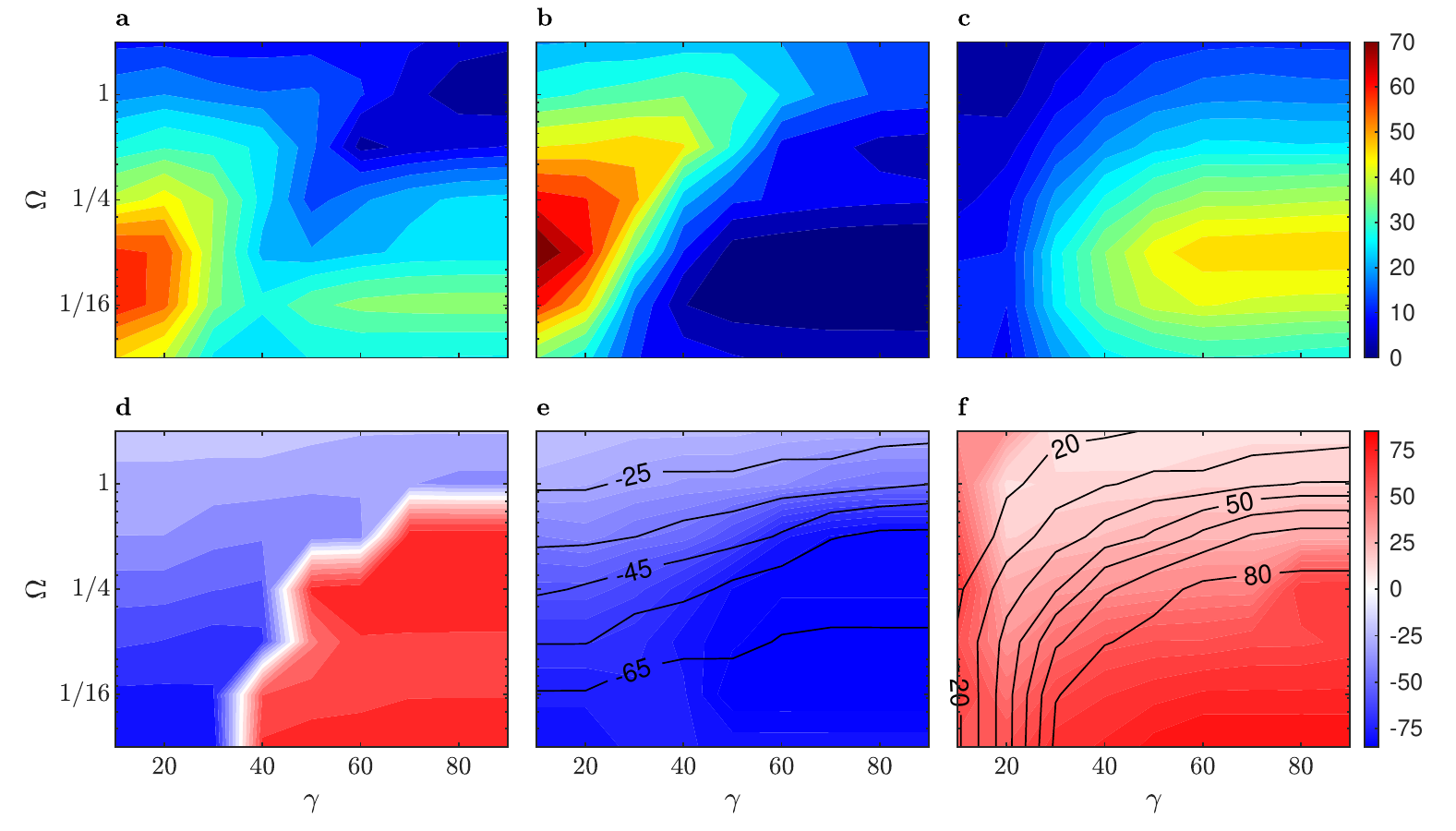}
    \caption{Global maximum wind (m s$^{-1}$, a) and its latitude (d) in the mid-troposphere ($\sigma=0.5$). High-level ($\sigma=0.25$) winter (southern) hemisphere maximum wind (m s$^{-1}$, b), its latitude (e, shading) and latitude of the Hadley cell descending branch (e, contours). Low-level ($\sigma=0.85$) summer (northern) hemisphere maximum wind (m s$^{-1}$, c), its latitude (f, shading) and latitude of the Hadley cell ascending branch (f, contours).}
    \label{fig:max_summer}
\end{figure*}

We find that westerly jets occur across the entire parameter space explored, with variability in their characteristics (Fig.~\ref{fig:jet5x5}). The most dramatic variation is a transition that occurs for high obliquity planets with a slow rotation rate, where the strongest jet is no longer in the winter hemisphere (like in the solar system's terrestrial planets), but rather in the mid-troposphere of the summer hemisphere (Figs. \ref{fig:jet5x5} and \ref{fig:max_summer}d). A closer examination of this transition reveals a gradual shift of low-level summer winds, which are observed in Earth's monsoon region and on Mars, towards the summer pole and their extension upward, towards the mid-troposphere (right column in Fig. \ref{fig:jet5x5}).

Both the winter jet and the mid-troposphere summer jet are strongly related to the cross-equatorial winter Hadley cell (contours in Fig.~\ref{fig:jet5x5}). While the winter jet is in close proximity to the descending branch of the cell, the summer jet is close to the ascending branch (Fig.~\ref{fig:max_summer}e-f). Their vertical structure also differs: The winter jet is near the top of the troposphere, extending downwards, whereas the summer jet is in the low and mid-troposphere (Fig.~\ref{fig:jet5x5}). Additionally, both the winter and summer jets' strength vary non-monotonically in the parameter space (Fig.~\ref{fig:max_summer}b-c).

To understand the flow's dependence on the obliquity and the rotation rate, we examine the parameter space starting with the low obliquity and a fast rotation rate case ($\gamma=10^{\circ}$ and $\Omega=1$, top left corner in Fig.~\ref{fig:jet5x5}). The flow, in this case, is close to hemispheric symmetry, with a slightly stronger jet in the winter hemisphere. As the rotation rate decreases (going down the first column in Fig.~\ref{fig:jet5x5}), the flow remains approximately hemispherically symmetric, with the circulation becoming wider and stronger and the jet strength varying non-monotonically with decreasing rotation rate (Figs.~\ref{fig:jet5x5} and \ref{fig:max_summer}a-b). This non-monotonic behavior was also noticed in previous studies \citep[e.g.,][]{Kaspi_2015, wang2018}. \cite{wang2018} argued that the non-monotonic response of the jet's strength to the reduction in the rotation rate is a result of a competing effect that occurs as the rotation rate decreases. On the one hand, as the rotation rate decreases, the jet's location shifts to a more poleward latitude and thus obtains more angular momentum. On the other hand, the reduction in the rotation rate reduces the global angular momentum, thereby diminishing the effect of the jet's shifting to a more poleward latitude, especially for slow rotation rates ($\Omega\leq1/8$), when the jet reaches high latitudes and no longer shifts poleward with reducing rotation rate, as it is close to the pole. 

In cases with low obliquities (leftmost column in Fig.~\ref{fig:jet5x5}), in addition to the winter jet, there is a jet in the summer hemisphere. Unlike the low-level summer jet discussed earlier, this summer hemisphere jet has similar characteristics to the winter jet, as it is a subtropical jet located at the descending edge of the summer Hadley circulation, unlike the low-level jet that is located at the ascending edge of the cell (Figs.~\ref{fig:jet5x5} and \ref{fig:max_summer}f). In addition to the widening of the circulation with decreasing rotation rate, equatorial superrotation also starts to develop at slow rotation rates \citep[Fig.~\ref{fig:jet5x5}, a feature seen on Venus and Titan, e.g.,][]{lebonnois_2014, Sanchez-Lavega2017}.

Increasing the obliquity for fast rotation rates (top row in Fig.~\ref{fig:jet5x5}, from left to right) increases the hemispherical asymmetry and pressure gradients, and the circulation consists of one dominant cross-equatorial winter cell. For fast rotation rates and high obliquities, the cross-equatorial Hadley circulation does not span the entire planet due to angular momentum constraints \citep{faulk2017, guendelman_2018,singh2019,hill2019}. These constraints are relaxed when the rotation rate is decreased, and, as a result, for intermediate-to-high obliquities, the circulation widens and the winter jet is diverted poleward as the rotation rate decreases (Fig.~\ref{fig:max_summer}c-e). 

As the mean meridional temperature gradients during winter increase with increasing obliquity \citep[e.g.,][]{Guendelman_2019, LOBO2020} one would expect a stronger jet with increasing obliquity. However, similar to the variations with rotation rate for the low obliquity case, the jet strength varies non-monotonically with the obliquity, reaching a maximum strength at low or moderate obliquities. Similar non-monotonic behaviour with obliquity was shown in Mars paleoclimate simulations \citep[e.g.,][]{TOIGO2020}. Angular momentum considerations again provide insights into the variations of jet strength in the parameter space. Consider the angular momentum conserving wind of a parcel at the jet latitude ($\phi_j$) that starts at rest at the ascending branch of the Hadley circulation ($\phi_1$)

\begin{equation}\label{eq:um}
    U_M = \Omega a \frac{\cos^2\phi_1-\cos^2\phi_j}{\cos\phi_j},
\end{equation} 
where $a$ is the planetary radius \citep[e.g.,][]{vallis_2017}. $U_M$ depends on whether $|\phi_j|$ is larger or smaller than $|\phi_1|$ and on how poleward $\phi_j$. For low obliquities, $\phi_1$ is close to the equator and $|\phi_j|-|\phi_1|$ is significantly greater than zero, so the jet is strong. As the obliquity increases, there are again competing effects, where on the one hand, $|\phi_j|-|\phi_1|$ becomes smaller and thus acting to reduce the jet strength, but on the other hand, $\phi_j$ moves more poleward, acting to increase the jet strength. For high obliquity, $|\phi_j|\sim|\phi_1|$ (Fig.~\ref{fig:max_summer}e,f) and the first effect dominates so the jet weakens. It is important to note that this is a very idealized representation, and in reality, there are other processes (e.g., eddies) that play a role in determining the jet's strength, however, it gives insight into the variation of the jet strength with obliquity.

The latitude with the maximum winds in the mid-troposphere ($\sigma=0.5$) changes sharply between hemispheres at high obliquities ($\gamma\ge 40$) and slow rotation rates ($\Omega<0.5$, Fig.~\ref{fig:max_summer}d), marking the transition from a winter jet to a summer jet regime. This regime transition occurs as the low-level winds in the summer hemisphere shift poleward and extend deeper in the atmosphere with increasing obliquity and decreasing rotation rate (Fig.~\ref{fig:jet5x5}). At moderate-to-high obliquities and fast rotation rates, the summer jet appears as low-level westerly winds close to the ascending edge of the Hadley cell (Fig.~\ref{fig:jet5x5},\ref{fig:max_summer}f). These low-level winds resemble the low-level westerlies that are observed on both Earth \citep[monsoon regions, e.g.,][]{joseph1966existence}, and Mars \citep[e.g.,][]{hinson1999} and are seen in Mars's paleoclimate studies \citep{HABERLE2003, TOIGO2020}. In these cases, during the solstice season, when the warmest latitude is off the equator, the circulation consists of a cross-equatorial circulation with air ascending in the summer hemisphere. The low-level returning poleward flow in the summer hemisphere is balanced by the boundary layer drag that results in the lower level westerlies \citep[e.g.,][]{schneider_bordoni_2008}.  

Across the parameter space, the low-level winds strengthen (non-monotonically with $\Omega$) and extend deeper into the mid-troposphere as the rotation rate decreases (Figs.~\ref{fig:jet5x5}-\ref{fig:max_summer}). The boundary layer process that explains the low-level winds in the fast rotating planets cannot explain the deepening of the jet at slow rotation rates. In the following section, we study the momentum balance to understand the jet changes across the parameter space.

\section{Balance and mechanism}\label{sec:balance}
\subsection{Meridional momentum balance}

\begin{figure*}[htb!]
    \centering
    \includegraphics[width=16cm]{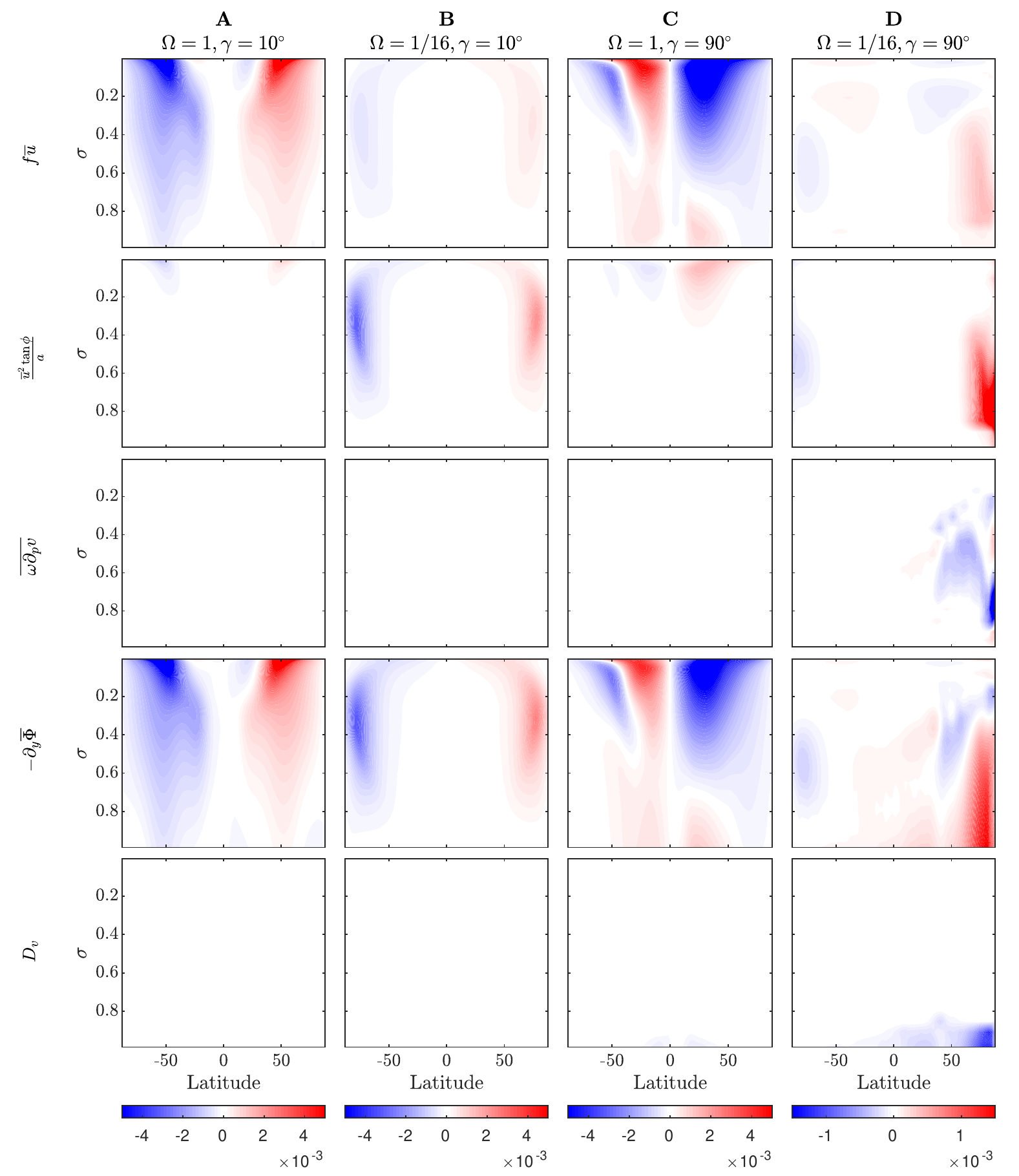}
    \caption{The different terms in the meridional momentum equation (rows, as written in Eq.~\ref{eq:meridional}, ms$^{-2}$) for four different cases (columns). Note different scales for different columns.}
    \label{fig:meridional}
\end{figure*}

In order to better understand the summer jet, it is insightful to consider the momentum balance. Assuming a steady-state, the zonal mean meridional momentum equation, in pressure coordinates, is given by \citep{vallis_2017}
\begin{equation}\label{eq:meridional}
    f\overline{u}+\frac{\overline{u^2}\tan\phi}{a}+\overline{\omega\frac{\partial v}{\partial p}} = -\frac{1}{a}\frac{\partial \overline{\Phi}}{\partial \phi} + D_v,
\end{equation}
where $v$ is the meridional velocity, $u$ the zonal velocity, $\omega$ the vertical (pressure) velocity, $f=2\Omega\sin\phi$ the Coriolis parameter, $\phi$ latitude, $\Phi$ the geopotential and $D_v$ the zonal mean meridional boundary layer drag. We neglect here merdional advection terms (terms of the form $\partial_{\phi}(v^2)$) and other forms of dissipation that are small, especially in the jet region. The over-bar notation denotes a zonal mean. Note, the steady-state assumption neglects the temporal variation in the momentum balance. This is a good approximation as the tendency term is two orders of magnitude smaller than the retained terms, even for strong seasonal cycles.

The planetary rotation rate influences the transition of the leading order meridional momentum balance (Eq.~\ref{eq:meridional}). For fast-rotating planets (Fig.~\ref{fig:meridional}A,C), such as Earth and Mars, the leading order balance is geostrophic \citep{vallis_2017}
\begin{equation}\label{eq:geostrophic}
    f\overline{u}\approx-\frac{1}{a}\frac{\partial \overline{\Phi}}{\partial \phi}.
\end{equation}
At moderate rotation rates, the leading order balance shifts to a thermal gradient balance \citep{sanchez2010introduction}  
\begin{equation}\label{eq:gradient}
    f\overline{u}+\frac{\overline{u}^2\tan\phi}{a} \approx -\frac{1}{a}\frac{\partial \overline{\Phi}}{\partial \phi}.
\end{equation}
For slow-rotating planets with low-to-moderate obliquity (Fig.~\ref{fig:meridional}B), such as Venus and Titan, the leading order balance becomes cyclostrophic \citep{read_2018}
\begin{equation}\label{eq:cyclostrophic}
    \frac{\overline{u}^2\tan\phi}{a} \approx -\frac{1}{a}\frac{\partial \overline{\Phi}}{\partial \phi}.
\end{equation}
The cyclostrophic balance does not hold for the summer jet, that occurs at slow rotation rates and high obliquities. Unlike all the above cases, where the vertical advection and friction are negligible, for slow rotation rates and high obliquities the leading order balance is more complex and all the terms in Eq.~\ref{eq:meridional} contribute to the balance (Fig.~\ref{fig:meridional}D). 

We can divide the summer jet balance into a boundary layer balance
\begin{equation}\label{eq:boundary}
    0 \approx -\frac{1}{a}\frac{\partial \overline{\Phi}}{\partial \overline{\phi}} + D_v,
\end{equation}
and a free atmosphere balance
\begin{equation}\label{eq:free}
    f\overline{u}+\frac{\overline{u^2}\tan\phi}{a}+\overline{\omega\frac{\partial v}{\partial p}} \approx -\frac{1}{a}\frac{\partial \overline{\Phi}}{\partial \phi}.
\end{equation}

The meridional momentum balance in both the boundary and free atmosphere differ between the summer jet and other regimes. First, for slow rotation rates and high obliquities, the boundary layer balance is comparable to the free atmosphere, but in all other cases, it is much weaker (Fig.~\ref{fig:meridional}). This is a result of the widening and strengthening of the circulation with decreasing rotation rate at high obliquity. As a result of the circulation becoming wider and stronger, the boundary layer poleward flow strengthens, and so does the drag that acts on it. Additionally, this effect is amplified by the weakening of the other terms (note the different scales in Fig.~\ref{fig:meridional}).

A second difference is in the boundary layer balance. In previous studies that explored either rapid rotation rates and moderate-high obliquity or moderate obliquity with slower rotation rates, the boundary layer balance was found to consist of a three- or four-way balance between the Coriolis acceleration, geopotential meridional gradient,  boundary layer drag, and, for slow rotation, the cyclostrophic term \citep[e.g.,][see also Fig.~\ref{fig:meridional}C]{schneider_bordoni_2008, faulk2017, LOBO2020}. However, for slow rotation rates and high obliquities, the boundary layer leading order balance is mainly between the drag and geopotential meridional gradient (Fig.~\ref{fig:meridional}D). Third, for slow rotation rates and high obliquities, the vertical advection term is essential, as it transports momentum between the boundary layer and the free atmosphere.

\begin{figure*}[!htb!]
    \centering
    \includegraphics[width=16cm]{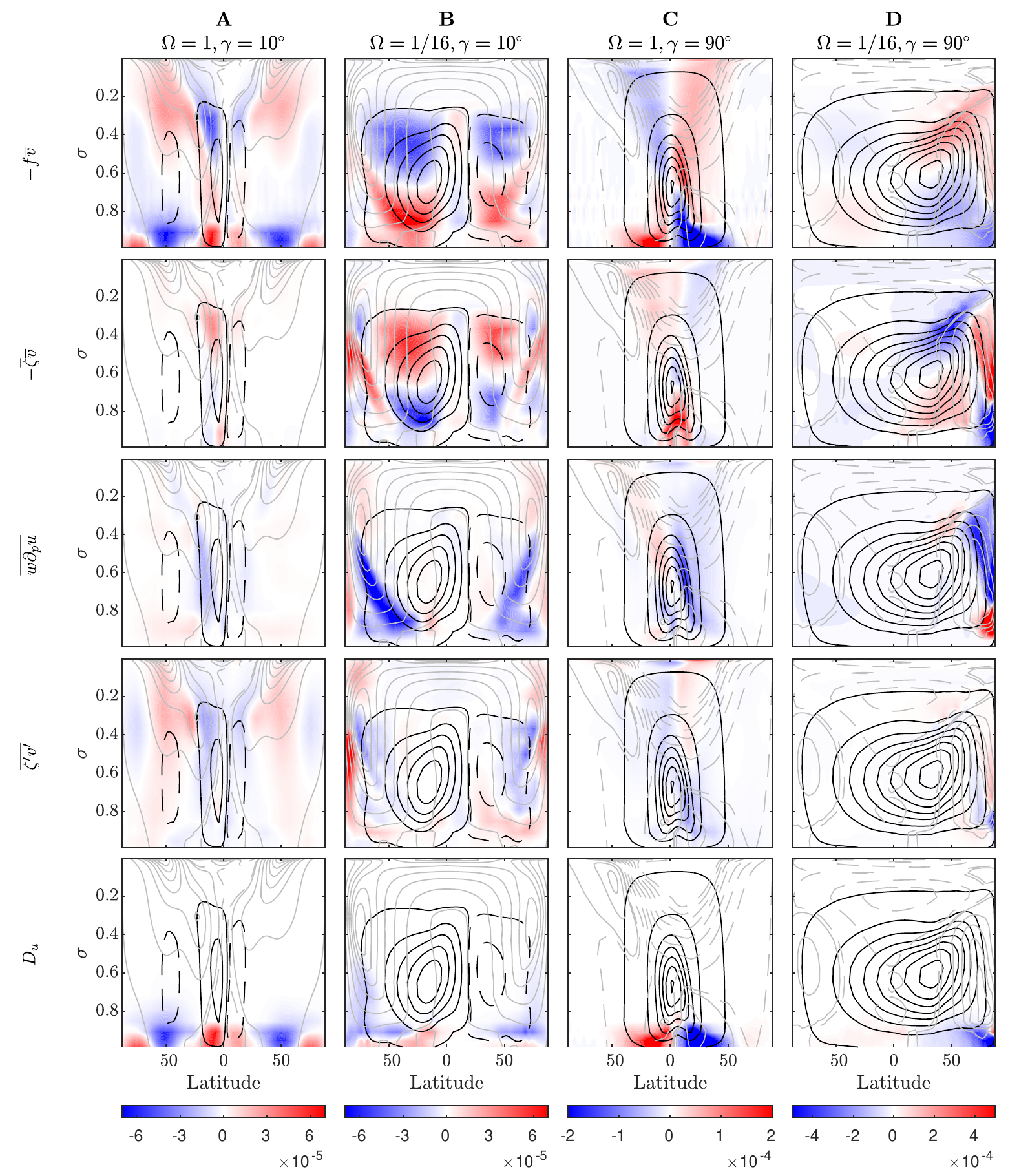}
    \caption{The different terms in the zonal momentum equation (rows, as written in Eq.~\ref{eq:zonal1}, ms$^{-2}$) for four different cases (columns). Black contours are for the mean meridional streamfunction, solid contours represent counterclockwise circulation. Contour intervals for fast rotation rate (A,C) as $1.5\times 10^{11}$ kg s$^{-1}$ and for slow rotation rate (B,D) $3\times 10^{11}$ kg s$^{-1}$. Gray contours are for the zonal mean zonal wind, contour interval is $10$ ms$^{-1}$. Note different scales for different columns.}
    \label{fig:zonal}
\end{figure*}

While the geostrophic and cyclostrophic balances relate the zonal wind to geopotential gradients, making the balance more intuitive, in the summer jet, the balance involves terms like vertical advection of meridional momentum and drag on meridional momentum, which makes the balance more complex and less straightforward to understand. To obtain a more complete picture of the summer jet's maintenance, it is important to understand the coupling between the meridional and zonal momentum balances. To do so, it is essential to understand the zonal mean zonal momentum balance. 

\subsection{Zonal momentum balance}

Assuming a steady state,  the zonal mean zonal momentum equation is given by \citep{vallis_2017}
\begin{equation}\label{eq:zonal1}
    -(f+\overline{\zeta})\overline{v}+\overline{\omega\frac{\partial u}{\partial p}} = \overline{\zeta'v'}+D_u.
\end{equation}
Eq.~\ref{eq:zonal1} is written using the vorticity notation, where $\overline{\zeta} =-(a\cos \phi)^{-1} \partial_{\phi}(\overline{u}\cos\phi)$, and $D_u$ is the zonal mean zonal boundary layer drag.

This balance gives insight into the processes responsible for the jet acceleration, as the terms of this balance represent zonal winds' tendencies. In general, when discussing the zonal mean circulation, one can distinguish between a flow mediated by eddies and one that is more axisymmetric. When the flow is axisymmetric, an air parcel at the top of the circulation moving poleward conserves its angular momentum, but this is not the case when eddies influence the circulation \citep[e.g.,][]{schneider_bordoni_2008}.

For fast rotation rates and low obliquities, the leading order balance at the top of the Hadley cell is $-(f+\overline{\zeta})\overline{v}\approx \overline{\zeta'v'}$, meaning that the circulation is influenced by eddies (Fig.~\ref{fig:zonal}A). Therefore, the angular momentum conserving wind is not a good approximation for the jet at the descending edge of the Hadley circulation \citep[e.g.,][]{schneider_2006}. When the rotation rate is decreased, while preserving a low-to-moderate obliquity or the obliquity is increased in a fast rotation rate scenario, there is a transition from an eddy-influenced circulation to a more axisymmetric one. For example, in slow rotation rates and a low obliquities, the leading order balance at the top of the circulation, except at the ascending and descending branches, is $-(f+\overline{\zeta})\overline{v}\approx 0$ (Fig.~\ref{fig:zonal}B). This balance means that the circulation in these regions is close to axisymmetric, and the jet is a result of angular momentum conservation at the top of the circulation \citep{held1980, lindzen_1988, schneider_bordoni_2008}. A similar balance appears for fast rotation rates with high obliquities, with the distinction in the boundary, where the boundary layer drag is stronger (Fig.~\ref{fig:zonal}C). This transition from an eddy-mediated circulation to a more axisymmetric one was previously shown for both increasing seasonality \citep[e.g.,][]{schneider_bordoni_2008, Bordoni2008, faulk2017, guendelman_2018, Guendelman_2019, LOBO2020, Guendelman_2020} and for decreasing rotation rate \citep[e.g.,][]{genio1987, Kaspi_2015, guendelman_2018, Guendelman_2019, coyler2019, Komacek_2019}.

In cases with slow rotation rates and high obliquities, away from the region of the summer jet, the balance is $-(f+\overline{\zeta})\overline{v}\approx 0$, meaning that the flow is close to axisymmetric. However, this is not the case in the summer jet, where the balance can be again divided into a boundary layer and a free atmosphere balance. Additionally, the balance differs between the regions poleward and equatorward of the jet core. Equatorward of the jet core, in the boundary layer the balance is $-(f+\overline{\zeta}) \overline{v}\approx D_u$, and at the top of the atmosphere, $(f+\overline{\zeta}) \overline{v}\approx \overline{\omega\partial_p u}$. Poleward of the jet core, the balance is somewhat different, and the boundary layer balance is $-(f+\overline{\zeta})\overline{v} + \overline{\omega\partial_p u}\approx D_u$, meaning that the returning flow of the Hadley circulation results in surface westerlies through an Ekman balance. The free atmosphere balance is $(f+\overline{\zeta}) \overline{v}\approx \overline{\omega\partial_p u}$, meaning that the mean meridional circulation is balanced by vertical advection of zonal momentum. Alternatively, momentum is transported vertically from the boundary layer to the free atmosphere (Fig.~\ref{fig:zonal}D).

\begin{figure}[htb!]
    \centering
    \includegraphics[width=7.5cm]{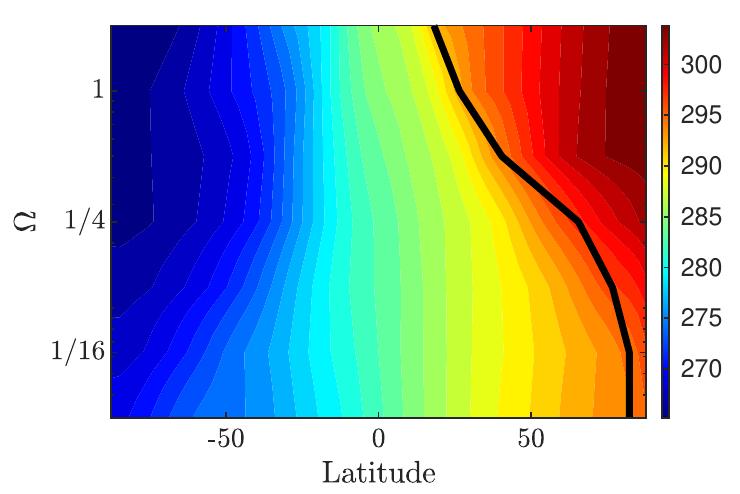}
    \caption{Near surface temperature for simulations with obliquity $50^{\circ}$ and different rotation rates (shadings, K) and the latitude of the Hadley cell ascending branch (black line).}
    \label{fig:temp_asc}
\end{figure}

For the summer jet to extend deeper into the mid-troposphere, the combination of high obliquity and a slow rotation rate is essential. Although past studies have connected the latitude of maximum moist static energy (and temperature) with the ascending motion at the edge of the Hadley cell \citep[e.g.,][]{prive2007}, recent studies have shown that this is not necessarily the case \citep[e.g.,][Fig.~\ref{fig:temp_asc}]{faulk2017, guendelman_2018, Guendelman_2019, LOBO2020}. This separation is a result of angular momentum constraints, and as the rotation rate decreases, the ascending motion shifts more towards the summer pole, i.e., closer to the latitude of maximum temperature \citep[e.g.,][]{guendelman_2018}. 

As the obliquity is increased, the latitude of maximum temperature moves poleward \citep[e.g.,][]{Guendelman_2019}, resulting in strong cross-equatorial circulation. Low-level westerlies balance the returning flow in the boundary layer. For fast rotation rates, the circulation width is confined, resulting in a separation between the warmest latitude and the Hadley cell ascending branch \citep[e.g.,][]{guendelman_2018, hill2019, singh2019}. Decreasing the rotation rate, the ascending branch of the Hadley cell aligns with the warmest latitude \citep[e.g.,][]{guendelman_2018, Guendelman_2019}. This transition is illustrated in Fig~\ref{fig:temp_asc} which shows the ascending branch latitude of the Hadley circulation (black line) and near surface temperature (shading). The alignment between the ascending motion of the Hadley circulation and the warmest latitude, together with a general strengthening of the circulation with decreasing rotation rate and increasing obliquity \citep{Guendelman_2019}, intensifies the ascending motion of the ascending branch of the cross-equatorial cell. This intensification is strongest close to the pole, the warmest latitude. As a result, the vertical motion at the polar latitude advects momentum upwards (more significantly poleward of the jet core), resulting in the low-level jet expanding into the mid-troposphere.

The separation between the Hadley cell ascending branch and the warmest latitude also explains the summer jet's split in latitude for intermediate rotation rates. At intermediate rotation rates, although the ascending branch is at the midlatitudes, there is a weak poleward flow poleward of the Hadley cell ascending branch, which also results in a westerly flow through an Ekman balance. Additionally, in the warmest latitude, there is an ascending motion of air due to local convection \citep{LOBO2020}. The combination of a weak poleward flow with an ascending motion in polar latitudes in these cases results in a secondary peak of zonal winds close to the pole at intermediate rotation rates, with the two peaks merging at slow rotation rates (two rightmost columns in Fig.~\ref{fig:jet5x5}).

\subsection{Non-dimensional numbers}\label{sec:non_dim}

\begin{figure*}[htb!]
    \centering
    \includegraphics[width=16cm]{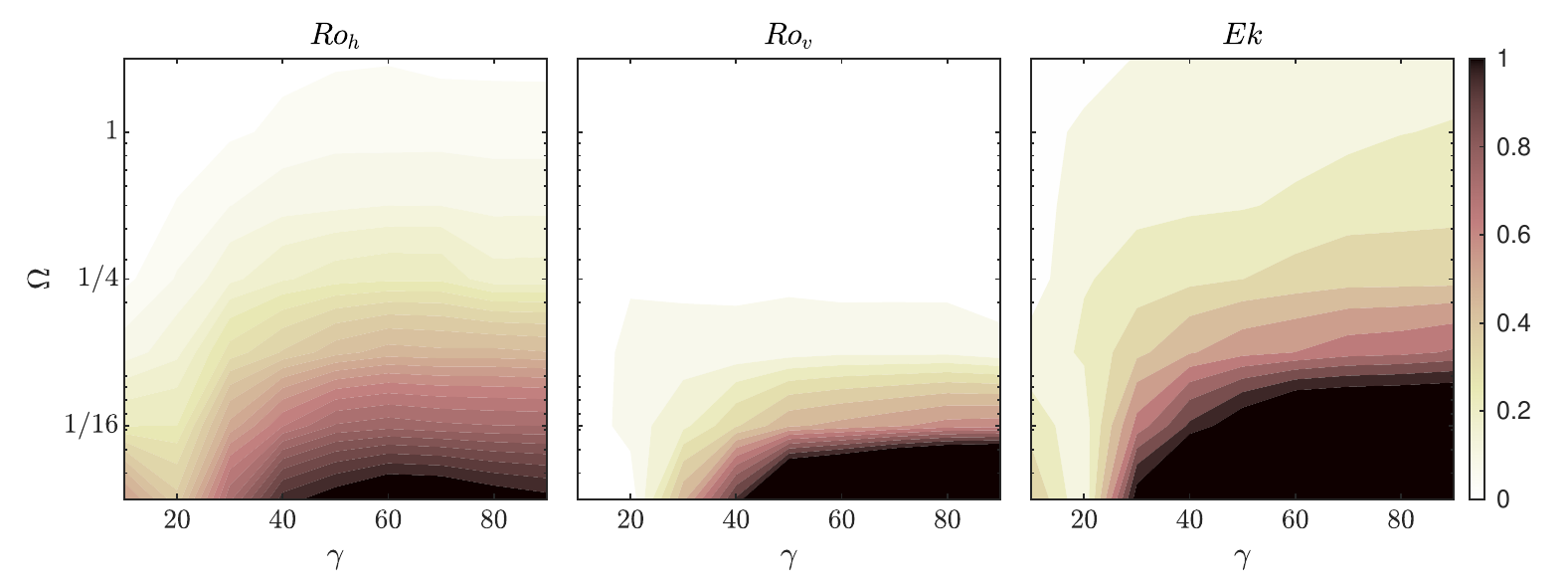}
    \caption{Values of non-dimensional numbers across the parameter space in the boundary layer in the summer jet core latitude. (a) $Ro_h$, horizontal Rossby number, (b) $Ro_v$, vertical Rossby number, and (c) $Ek$ Ekman number from the zonal mean meridional equation (Eq.~\ref{eq:nondim}).}
    \label{fig:non_dom}
\end{figure*}

An alternative way to understand the transition between the winter- and summer-jet regimes, is by considering non-dimensional numbers. Non-dimensionalizing the zonal mean meridional equation (Eq.~\ref{eq:meridional}) gives 

\begin{equation}\label{eq:nondim}
    \hat{u}+Ro_h\hat{u^2}\tan\phi+Ro_v\hat{\omega}\frac{\partial \hat{v}}{\partial \hat{p}} = -\frac{\partial \hat{\Phi}}{\partial \phi} + Ek,
\end{equation}
where we scale the horizontal wind as $u,v\sim U$, the vertical wind as $\omega\sim W$, and
\begin{equation}
    Ro_h=U/fa
\end{equation}
is the horizontal Rossby number,
\begin{equation}
    Ro_v =W /f\Delta p
\end{equation} 
is the vertical Rossby number,
\begin{equation}
    Ek = D_v/fu
\end{equation} 
is the Ekman number, and
\begin{equation}
    \hat{\Phi} = \frac{\Phi}{fUa},
\end{equation}
is the non-dimensional geopotential. The planetary radius, $a$, is taken as the typical horizontal length scale, and $\Delta p$ is taken as the typical height of the circulation (in pressure coordinates). For the winter jet, where the jet is in the free atmosphere, surface drag and vertical advection are negligible ($Ro_v, Ek\ll1$) and the transition from a geostrophic balance (Eq.~\ref{eq:geostrophic}) to a thermal grandient balance (Eq.~\ref{eq:gradient}) and then a cyclostrophic balance (Eq.~\ref{eq:cyclostrophic}) is represented by the increase in $Ro_h$ (Fig.~\ref{fig:non_dom}).

The variations of the non-dimensional numbers in the boundary layer indicate that the transition to the summer jet regime occurs when all three non-dimensional parameters are of order one or larger (Fig.~\ref{fig:non_dom}). Although all the non-dimensional parameters increase with decreasing rotation rate and increasing obliquity, the increase in $Ro_v$ at low rotation rates and moderate-high obliquities is the most abrupt. This represents the transition from a low-level summer jet to a mid-tropospheric jet, due to the intensification of the vertical advection as the Hadley cell's ascending branch and warmest latitude align.

In addition, the non-dimensional numbers vary spatially and can represent the different balance between the boundary layer and free atmosphere. For example, in the summer jet regime, in the boundary layer, $Ro_h,Ro_v>1$, and the balance is given by Eq.~\ref{eq:boundary} (Fig.~\ref{fig:non_dom}), whereas in the free troposphere, $Ro_h,Ro_v\sim 1$ and $Ek\ll1$ and the balance is given by Eq.~\ref{eq:free}.  

\section{Conclusions}\label{sec:conc}

The solar system's terrestrial planetary bodies with a seasonal cycle have a dominant jet in the winter hemisphere, close to the Hadley cell descending branch. In the monsoon region on Earth, where the ascending motion of the circulation is most dominant \citep{raiter2020}, and on Mars, this winter jet is accompanied by weaker low-level westerlies close to the Hadley cell ascending branch in the summer hemisphere (Fig.~\ref{fig:obs}). To better understand this phenomenon and general jet dynamics, we examine the jet characteristics' dependence on the rotation rate and obliquity, which allows a better separation of scales needed to identify the physical mechanisms controlling the dynamics. Consistent with observations of the solar system, at a fast rotation rate and strong seasonality, the dominant jet is in the winter hemisphere. However, we show that for planets with high obliquity and a slow rotation rate, the low-level summer westerlies extend into the middle troposphere, and the dominant jet is in the summer hemisphere and occurs within the ascending, rather than descending, branch of the Hadley cell (Figs.~\ref{fig:jet5x5},\ref{fig:max_summer}).

\begin{figure*}[htb!]
    \centering
    \includegraphics[width=16cm]{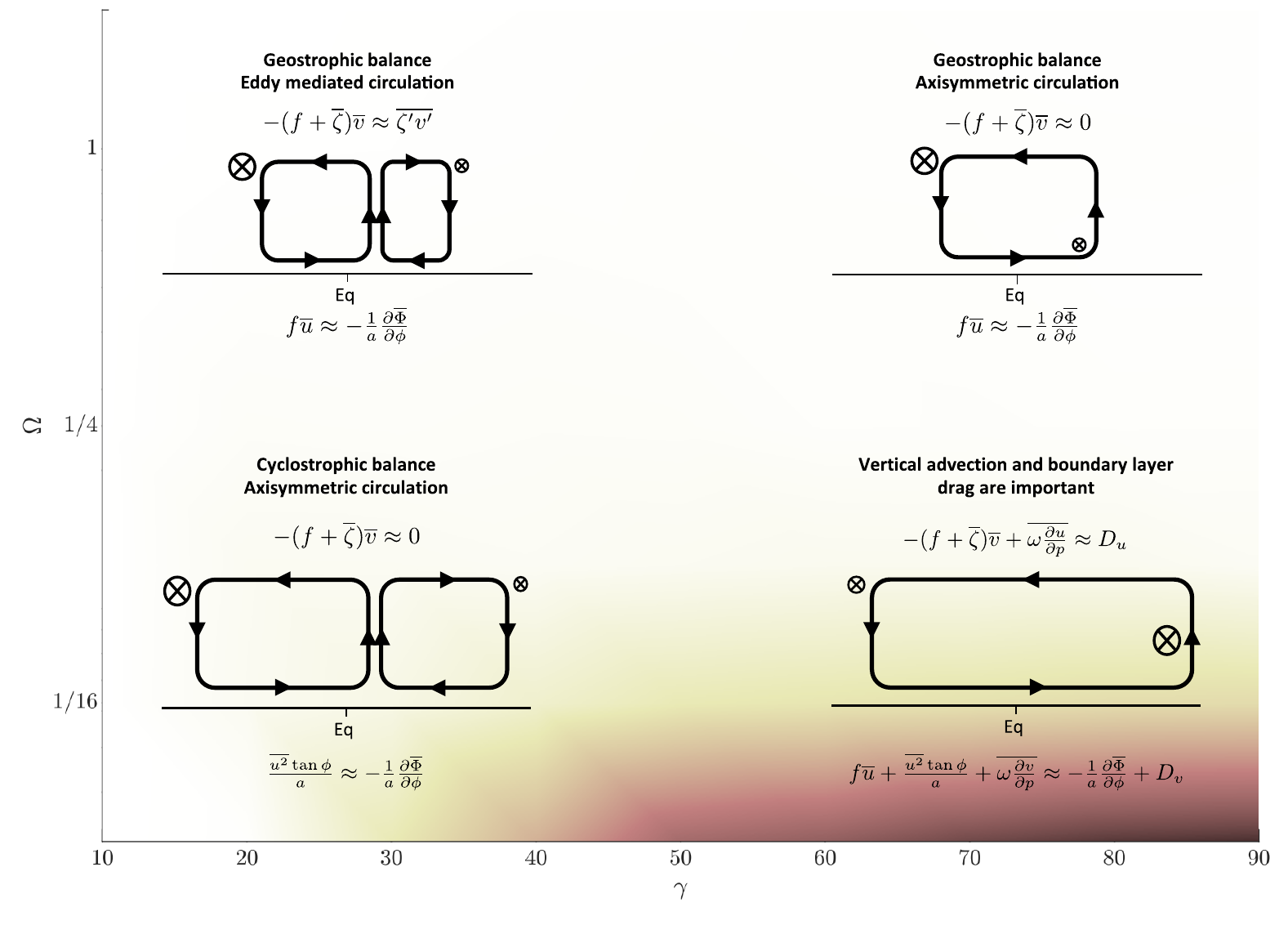}
    \caption{Schematic picture summarizing the different circulation regimes presented in this study. The background shading shows the value of $Ro_v=\omega/f\Delta p$, with the scale between $0$ (white) and $2$ (dark).}
    \label{fig:schematic}
\end{figure*}

Figure \ref{fig:schematic} summarises schematically the different circulation regimes that exist in this parameter space. At fast rotation rates and low obliquities, the mean meridional circulation is strongly influenced by eddies and consists of two cells (Fig.~\ref{fig:zonal}A), with the cross-equatorial winter cell becoming wider and stronger. At the edge of each cell (winter and summer), there is a westerly jet that is in a geostrophic balance (Fig.~\ref{fig:meridional}A), with the winter hemisphere jet being stronger than the summer jet. For slower rotation rates and low obliquities, the circulation still consists of two cells that are wider and more axisymmetric (Fig.~\ref{fig:zonal}B). The jets at the descending branch of each cell are now in a cyclostrophic balance (Fig.~\ref{fig:meridional}B), with the winter jet the more dominant one.

Once the obliquity is high enough, the circulation consists of one strong cross-equatorial cell that becomes more axisymmetric. The returning boundary layer poleward flow in the summer hemisphere is balanced by a low-level westerly flow through an Ekman balance (Fig.~\ref{fig:zonal}C). For fast rotation rates, the circulation's width is limited and does not align with the warmest latitude due to angular momentum constraints \citep{guendelman_2018, hill2019, singh2019}. In this case, the summer hemisphere westerlies remain close to the boundary layer, and the winter jet is the dominant jet. As the rotation rate is decreased, the Hadley cell ascending branch and the warmest latitude align (Fig.~\ref{fig:temp_asc}). This alignment, together with the strengthening of the circulation with increasing obliquity and decreasing rotation rate \citep{Guendelman_2019}, results in a more efficient vertical advection of momentum, extending the winds further upwards in the atmosphere, which results in a mid-troposphere summer jet, which also becomes the more dominant jet (Fig.~\ref{fig:zonal}D). 

The importance of the vertical advection in the momentum balance can be represented by the increase in the vertical Rossby number, $Ro_v$. For moderate-fast rotation rates or moderate-low obliquities, $Ro_v\ll 1$ and the vertical advection is negligible. However, for slow rotation rates and high obliquities, $Ro_v\sim 1$ and the vertical advection cannot be neglected (shading in Fig.~\ref{fig:schematic}). 

The results presented here increase our understanding of the wind patterns on terrestrial planets in the solar system and provide insights into possible flows on exoplanets. While a strong winter westerly jet is a common feature of terrestrial planets in the solar system, the simulations presented here indicate this may not always be the case for all exoplanets. In particular, if a terrestrial exoplanet has a slow rotation rate like Venus and Titan but a higher obliquity, it may have summer, rather than winter, westerly jets. The impact of a summer jet needs to be further examined. For example, one possible impact of a summer jet is an effect on the seasonality of trace constituents transport to polar regions. On Earth, Mars, and Titan, the winter jets act to reduce transport between low and high latitudes, and unique chemical-microphysical processes occur in these planets' winter polar atmospheres. The situation may be different for planets with dominant summer jets.

%








%
\acknowledgments
We thank Janny Yuval, Dorian Abbot and two anonymous reviewers for helpful comments. We acknowledge support from the Minerva Foundation and the Helen Kimmel Center of Planetary Science at the Weizmann Institute of Science. 
%






%
%
%
\bibliographystyle{ametsoc2014}
\bibliography{jets.bib}

\begin{thebibliography}{46}
\providecommand{\natexlab}[1]{#1}
\providecommand{\url}[1]{\texttt{#1}}
\renewcommand{\UrlFont}{\rmfamily}
\providecommand{\urlprefix}{URL }
\expandafter\ifx\csname urlstyle\endcsname\relax
  \providecommand{\doi}[1]{doi:\discretionary{}{}{}#1}\else
  \providecommand{\doi}{doi:\discretionary{}{}{}\begingroup
  \urlstyle{rm}\Url}\fi
\providecommand{\eprint}[2][]{\url{#2}}

\bibitem[{Bordoni and Schneider(2008)Bordoni, and Schneider}]{Bordoni2008}
Bordoni, S., and T.~Schneider, 2008: Monsoons as eddy-mediated regime
  transitions of the tropical overturning circulation. \textit{Nature
  Geoscience}, \textbf{1~(8)}, 515--519, \doi{10.1038/ngeo248}.

\bibitem[{Colyer and Vallis(2019)Colyer, and Vallis}]{coyler2019}
Colyer, G.~J., and G.~K. Vallis, 2019: Zonal-mean atmospheric dynamics of
  slowly rotating terrestrial planets. \textit{J. Atmos. Sci.},
  \textbf{76~(5)}, 1397--1418, \doi{10.1175/JAS-D-18-0180.1}.

\bibitem[{de~Kok et~al.(2014)de~Kok, Teanby, Maltagliati, Irwin,, and
  Vinatier}]{DeKok2014}
de~Kok, R.~J., N.~A. Teanby, L.~Maltagliati, P.~G.~J. Irwin, and S.~Vinatier,
  2014: {HCN} ice in {T}itan's high-altitude southern polar cloud.
  \textit{Nature}, \textbf{514~(7520)}, 65--67, \doi{10.1038/nature13789}.

\bibitem[{Del~Genio and Suozzo(1987)Del~Genio, and Suozzo}]{genio1987}
Del~Genio, A.~D., and R.~J. Suozzo, 1987: A comparative study of rapidly and
  slowly rotating dynamical regimes in a terrestrial general circulation model.
  \textit{J. Atmos. Sci.}, \textbf{44~(6)}, 973--986,
  \doi{10.1175/1520-0469(1987)044<0973:ACSORA>2.0.CO;2}.

\bibitem[{Faulk et~al.(2017)Faulk, Mitchell,, and Bordoni}]{faulk2017}
Faulk, S., J.~Mitchell, and S.~Bordoni, 2017: Effects of rotation rate and
  seasonal forcing on the itcz extent in planetary atmospheres. \textit{J.
  Atmos. Sci.}, \textbf{74~(3)}, 665--678, \doi{10.1175/JAS-D-16-0014.1}.

\bibitem[{Ferreira et~al.(2014)Ferreira, Marshall, O’Gorman,, and
  Seager}]{FERREIRA2014}
Ferreira, D., J.~Marshall, P.~A. O’Gorman, and S.~Seager, 2014: Climate at
  high-obliquity. \textit{Icarus}, \textbf{243}, 236 -- 248,
  \doi{10.1016/j.icarus.2014.09.015}.

\bibitem[{Findlater(1969)}]{findlater1969}
Findlater, J., 1969: A major low-level air current near the indian ocean during
  the northern summer. \textit{Q. J. R. Meteorol. Soc.}, \textbf{95~(404)},
  362--380, \doi{https://doi.org/10.1002/qj.49709540409}.

\bibitem[{Flasar and Achterberg(2009)Flasar, and Achterberg}]{flasar2009}
Flasar, F., and R.~Achterberg, 2009: The structure and dynamics of {T}itan's
  middle atmosphere. \textit{Philos. Trans. Roy. Soc. A}, \textbf{367~(1889)},
  649--664, \doi{10.1098/rsta.2008.0242}.

\bibitem[{Frierson et~al.(2006)Frierson, Held,, and
  Zurita-Gotor}]{frierson2006}
Frierson, D. M.~W., I.~M. Held, and P.~Zurita-Gotor, 2006: A gray-radiation
  aquaplanet moist gcm. part {I}: Static stability and eddy scale. \textit{J.
  Atmos. Sci.}, \textbf{63~(10)}, 2548--2566, \doi{10.1175/JAS3753.1}.

\bibitem[{Greybush et~al.(2019)}]{greybush2019}
Greybush, S.~J., and Coauthors, 2019: The ensemble mars atmosphere reanalysis
  system (emars) version 1.0. \textit{Geoscience Data Journal}, \textbf{6~(2)},
  137--150, \doi{https://doi.org/10.1002/gdj3.77}.

\bibitem[{Guendelman and Kaspi(2018)Guendelman, and Kaspi}]{guendelman_2018}
Guendelman, I., and Y.~Kaspi, 2018: An axisymmetric limit for the width of the
  hadley cell on planets with large obliquity and long seasonality.
  \textit{Geophys. Res. Lett.}, \textbf{45~(24)}, 13,213--13,221,
  \doi{10.1029/2018GL080752}.

\bibitem[{Guendelman and Kaspi(2019)Guendelman, and Kaspi}]{Guendelman_2019}
Guendelman, I., and Y.~Kaspi, 2019: Atmospheric dynamics on terrestrial
  planets: The seasonal response to changes in orbital, rotational, and
  radiative timescales. \textit{Astrophys. J.}, \textbf{881~(1)}, 67,
  \doi{10.3847/1538-4357/ab2a06}.

\bibitem[{Guendelman and Kaspi(2020)Guendelman, and Kaspi}]{Guendelman_2020}
Guendelman, I., and Y.~Kaspi, 2020: Atmospheric dynamics on terrestrial planets
  with eccentric orbits. \textit{Astrophys. J.}, \textbf{901~(1)}, 46,
  \doi{10.3847/1538-4357/abaef8}.

\bibitem[{Haberle et~al.(2003)Haberle, Murphy,, and Schaeffer}]{HABERLE2003}
Haberle, R.~M., J.~R. Murphy, and J.~Schaeffer, 2003: Orbital change
  experiments with a mars general circulation model. \textit{Icarus},
  \textbf{161~(1)}, 66 -- 89, \doi{10.1016/S0019-1035(02)00017-9}.

\bibitem[{Haberle et~al.(1993)Haberle, Pollack, Barnes, Zurek, Leovy, Murphy,
  Lee,, and Schaeffer}]{haberle_1993}
Haberle, R.~M., J.~B. Pollack, J.~R. Barnes, R.~W. Zurek, C.~B. Leovy, J.~R.
  Murphy, H.~Lee, and J.~Schaeffer, 1993: Mars atmospheric dynamics as
  simulated by the nasa ames general circulation model: 1. the zonal-mean
  circulation. \textit{J. Geophys. Res. (Planets)}, \textbf{98~(E2)},
  3093--3123, \doi{10.1029/92JE02946}.

\bibitem[{Hayes et~al.(2013)}]{HAYES2013}
Hayes, A., and Coauthors, 2013: Wind driven capillary-gravity waves on
  {T}itan’s lakes: Hard to detect or non-existent? \textit{Icarus},
  \textbf{225~(1)}, 403 -- 412, \doi{10.1016/j.icarus.2013.04.004}.

\bibitem[{Held and Hou(1980)Held, and Hou}]{held1980}
Held, I.~M., and A.~Y. Hou, 1980: Nonlinear axially symmetric circulations in a
  nearly inviscid atmosphere. \textit{J. Atmos. Sci.}, \textbf{37~(3)},
  515--533, \doi{10.1175/1520-0469(1980)037<0515:NASCIA>2.0.CO;2}.

\bibitem[{Hill et~al.(2019)Hill, Bordoni,, and Mitchell}]{hill2019}
Hill, S.~A., S.~Bordoni, and J.~L. Mitchell, 2019: Axisymmetric constraints on
  cross-equatorial hadley cell extent. \textit{J. Atmos. Sci.},
  \textbf{76~(6)}, 1547--1564, \doi{10.1175/JAS-D-18-0306.1}.

\bibitem[{Hinson et~al.(1999)Hinson, Simpson, Twicken, Tyler,, and
  Flasar}]{hinson1999}
Hinson, D.~P., R.~A. Simpson, J.~D. Twicken, G.~L. Tyler, and F.~M. Flasar,
  1999: Initial results from radio occultation measurements with {M}ars global
  surveyor. \textit{J. Geophys. Res. (Planets)}, \textbf{104~(E11)},
  26\,997--27\,012, \doi{10.1029/1999JE001069}.

\bibitem[{Hofgartner et~al.(2016)}]{HOFGARTNER2016}
Hofgartner, J.~D., and Coauthors, 2016: Titan’s “magic islands”:
  Transient features in a hydrocarbon sea. \textit{Icarus}, \textbf{271}, 338
  -- 349, \doi{10.1016/j.icarus.2016.02.022}.

\bibitem[{Joseph and Raman(1966)Joseph, and Raman}]{joseph1966existence}
Joseph, P.~V., and P.~L. Raman, 1966: Existence of low level westerly jet
  stream over peninsular india during {J}uly. \textit{Indian J. Meteorol.
  Geophys.}, \textbf{17~(1)}, 407--410.

\bibitem[{Kaspi and Showman(2015)Kaspi, and Showman}]{Kaspi_2015}
Kaspi, Y., and A.~P. Showman, 2015: Atmospheric dynamics of terrestrial
  exoplanets over a wide range of orbital and atmospheric parameters.
  \textit{Astrophys. J.}, \textbf{804~(1)}, 60,
  \doi{10.1088/0004-637x/804/1/60}.

\bibitem[{Komacek and Abbot(2019)Komacek, and Abbot}]{Komacek_2019}
Komacek, T.~D., and D.~S. Abbot, 2019: The atmospheric circulation and climate
  of terrestrial planets orbiting sun-like and m dwarf stars over a broad range
  of planetary parameters. \textit{Astrophys. J.}, \textbf{871~(2)}, 245,
  \doi{10.3847/1538-4357/aafb33}.

\bibitem[{Lebonnois et~al.(2014)Lebonnois, Flasar, Tokano,, and
  Newman}]{lebonnois_2014}
Lebonnois, S., F.~M. Flasar, T.~Tokano, and C.~E. Newman, 2014: The general
  circulation of {T}itan's lower and middle atmosphere. \textit{{Titan:
  Interior, Surface, Atmosphere and Space Environment}}, {Cambridge University
  Press}, 122--157.

\bibitem[{Lindzen and Hou(1988)Lindzen, and Hou}]{lindzen_1988}
Lindzen, R.~S., and A.~V. Hou, 1988: Hadley circulations for zonally averaged
  heating centered off the equator. \textit{J. Atmos. Sci.}, \textbf{45~(17)},
  2416--2427, \doi{10.1175/1520-0469(1988)045<2416:HCFZAH>2.0.CO;2}.

\bibitem[{Linsenmeier et~al.(2015)Linsenmeier, Pascale,, and
  Lucarini}]{LINSENMEIER2015}
Linsenmeier, M., S.~Pascale, and V.~Lucarini, 2015: Climate of earth-like
  planets with high obliquity and eccentric orbits: Implications for
  habitability conditions. \textit{Planatary and Space Science}, \textbf{105},
  43 -- 59, \doi{https://doi.org/10.1016/j.pss.2014.11.003}.

\bibitem[{Lobo and Bordoni(2020)Lobo, and Bordoni}]{LOBO2020}
Lobo, A.~H., and S.~Bordoni, 2020: Atmospheric dynamics in high obliquity
  planets. \textit{Icarus}, \textbf{340}, 113\,592,
  \doi{10.1016/j.icarus.2019.113592}.

\bibitem[{Mitchell et~al.(2014)Mitchell, Vallis,, and Potter}]{Mitchell_2014}
Mitchell, J.~L., G.~K. Vallis, and S.~F. Potter, 2014: Effects of the seasonal
  cycle on superrotation in planetary atmospheres. \textit{Astrophys. J.},
  \textbf{787~(1)}, 23, \doi{10.1088/0004-637x/787/1/23}.

\bibitem[{Privé and Plumb(2007)Privé, and Plumb}]{prive2007}
Privé, N.~C., and R.~A. Plumb, 2007: Monsoon dynamics with interactive
  forcing. part i: Axisymmetric studies. \textit{J. Atmos. Sci.},
  \textbf{64~(5)}, 1417 -- 1430, \doi{10.1175/JAS3916.1}.

\bibitem[{Raiter et~al.(2020)Raiter, Galanti,, and Kaspi}]{raiter2020}
Raiter, D., E.~Galanti, and Y.~Kaspi, 2020: The tropical atmospheric conveyor
  belt: A coupled eulerian-lagrangian analysis of the large-scale tropical
  circulation. \textit{Geophys. Res. Lett.}, \textbf{47~(10)}, e2019GL086\,437,
  \doi{https://doi.org/10.1029/2019GL086437}.

\bibitem[{Read and Lebonnois(2018)Read, and Lebonnois}]{read_2018}
Read, P.~L., and S.~Lebonnois, 2018: Superrotation on venus, on titan, and
  elsewhere. \textit{Ann. Rev. Earth Plan. Sci.}, \textbf{46~(1)}, 175--202,
  \doi{10.1146/annurev-earth-082517-010137}.

\bibitem[{Sanchez-Lavega(2010)}]{sanchez2010introduction}
Sanchez-Lavega, A., 2010: \textit{An introduction to planetary atmospheres}.
  Taylor \& Francis.

\bibitem[{S{\'{a}}nchez-Lavega et~al.(2017)S{\'{a}}nchez-Lavega, Lebonnois,
  Imamura, Read,, and Luz}]{Sanchez-Lavega2017}
S{\'{a}}nchez-Lavega, A., S.~Lebonnois, T.~Imamura, P.~Read, and D.~Luz, 2017:
  The atmospheric dynamics of {V}enus. \textit{Space Sci. Rev.},
  \textbf{212~(3-4)}, 1541--1616, \doi{10.1007/s11214-017-0389-x}.

\bibitem[{Schneider(2006)}]{schneider_2006}
Schneider, T., 2006: The general circulation of the atmosphere. \textit{Ann.
  Rev. Earth Plan. Sci.}, \textbf{34~(1)}, 655--688,
  \doi{10.1146/annurev.earth.34.031405.125144}.

\bibitem[{Schneider and Bordoni(2008)Schneider, and
  Bordoni}]{schneider_bordoni_2008}
Schneider, T., and S.~Bordoni, 2008: Eddy-mediated regime transitions in the
  seasonal cycle of a hadley circulation and implications for monsoon dynamics.
  \textit{J. Atmos. Sci.}, \textbf{65~(3)}, 915--934,
  \doi{10.1175/2007JAS2415.1}.

\bibitem[{Schoeberl and Hartmann(1991)Schoeberl, and Hartmann}]{schoeberl1991}
Schoeberl, M.~R., and D.~L. Hartmann, 1991: The dynamics of the stratospheric
  polar vortex and its relation to springtime ozone depletions.
  \textit{Science}, \textbf{251~(4989)}, 46--52,
  \doi{10.1126/science.251.4989.46}.

\bibitem[{Singh(2019)}]{singh2019}
Singh, M.~S., 2019: Limits on the extent of the solsticial hadley cell: The
  role of planetary rotation. \textit{J. Atmos. Sci.}, \textbf{76~(7)},
  1989--2004, \doi{10.1175/JAS-D-18-0341.1}.

\bibitem[{Tan et~al.(2019)Tan, Lachmy,, and Shaw}]{tan2019}
Tan, Z., O.~Lachmy, and T.~A. Shaw, 2019: The sensitivity of the jet stream
  response to climate change to radiative assumptions. \textit{J. Adv. Model.
  Earth Syst.}, \textbf{11~(4)}, 934--956,
  \doi{https://doi.org/10.1029/2018MS001492}.

\bibitem[{Toigo et~al.(2017)Toigo, Waugh,, and Guzewich}]{toigo2017}
Toigo, A.~D., D.~W. Waugh, and S.~D. Guzewich, 2017: What causes mars' annular
  polar vortices? \textit{Geophys. Res. Lett.}, \textbf{44~(1)}, 71--78,
  \doi{https://doi.org/10.1002/2016GL071857}.

\bibitem[{Toigo et~al.(2020)Toigo, Waugh,, and Guzewich}]{TOIGO2020}
Toigo, A.~D., D.~W. Waugh, and S.~D. Guzewich, 2020: Atmospheric transport into
  polar regions on mars in different orbital epochs. \textit{Icarus},
  \textbf{347}, 113\,816, \doi{10.1016/j.icarus.2020.113816}.

\bibitem[{Vallis(2017)}]{vallis_2017}
Vallis, G.~K., 2017: \textit{Atmospheric and Oceanic Fluid Dynamics:
  Fundamentals and Large-Scale Circulation}. 2nd ed., Cambridge University
  Press, \doi{10.1017/9781107588417}.

\bibitem[{Wang et~al.(2018)Wang, Read, Tabataba-Vakili,, and Young}]{wang2018}
Wang, Y., P.~L. Read, F.~Tabataba-Vakili, and R.~M.~B. Young, 2018: Comparative
  terrestrial atmospheric circulation regimes in simplified global circulation
  models. part {I}: From cyclostrophic super-rotation to geostrophic
  turbulence. \textit{Q. J. R. Meteorol. Soc.}, \textbf{144~(717)}, 2537--2557,
  \doi{10.1002/qj.3350}.

\bibitem[{Waugh et~al.(2017)Waugh, Sobel,, and Polvani}]{waugh2017}
Waugh, D.~W., A.~H. Sobel, and L.~M. Polvani, 2017: What is the polar vortex
  and how does it influence weather? \textit{Bull. Am. Meteor. Soc.},
  \textbf{98~(1)}, 37--44, \doi{10.1175/BAMS-D-15-00212.1}.

\bibitem[{Waugh et~al.(2016)Waugh, Toigo, Guzewich, Greybush, Wilson,, and
  Montabone}]{waugh_2016}
Waugh, D.~W., A.~D. Toigo, S.~D. Guzewich, S.~J. Greybush, R.~J. Wilson, and
  L.~Montabone, 2016: Martian polar vortices: Comparison of reanalyses.
  \textit{J. Geophys. Res. (Planets)}, \textbf{121~(9)}, 1770--1785,
  \doi{10.1002/2016JE005093}.

\bibitem[{Williams(1988)}]{Williams1988}
Williams, G.~P., 1988: The dynamical range of global circulations — ii.
  \textit{Clim. Dyn.}, \textbf{3~(2)}, 45--84, \doi{10.1007/BF01080901}.

\bibitem[{Williams and Holloway(1982)Williams, and Holloway}]{Williams1982}
Williams, G.~P., and J.~L. Holloway, 1982: The range and unity of planetary
  circulations. \textit{Nature}, \textbf{297~(5864)}, 295--299,
  \doi{10.1038/297295a0}.

\end{thebibliography}

%

%

\end{document}